\documentstyle[12pt,aaspp4]{article}

\def\linebreak{\hfil\break}

\def\
{\hfil\linebreak}
\def\singlespace{\baselineskip=14pt}

\def\singlespace{%
    \lineskip                .15ex
    \baselineskip            3.0ex
   \lineskiplimit              0ex
   \parskip                0.60ex plus .30ex minus .15ex
   }%

\def\etal{{\it et al}. }
\def\degree{\ifmmode {^\circ}\else {$^\circ$}\fi}
\def\mum{\ifmmode {\rm \mu {\rm m}}\else $\rm \mu {\rm m}$\fi}
\def\arcsec{\ifmmode ^{\prime \prime}\else $^{\prime \prime}$\fi}

\def\inch{\ifmmode ^{\prime \prime}\else $^{\prime \prime}$\fi}
\def\arcmin{\ifmmode ^{\prime}\else $^{\prime}$\fi}

\def\msun{\ifmmode {\rm M_{\odot}}\else $\rm M_{\odot}$\fi}
\def\mearth{\ifmmode {\rm M_{+\mskip-14.6muO\,}}\else $\rm M_{+\mskip-14.6muO\,}$\fi}
\def\mearth{\ifmmode {\rm M_{\earth}}\else $\rm M_{\earth}$\fi}
\newbox\grsign \setbox\grsign=\hbox{$>$} \newdimen\grdimen \grdimen=\ht\grsign
\newbox\simlessbox \newbox\simgreatbox
\setbox\simgreatbox=\hbox{\raise.5ex\hbox{$>$}\llap
     {\lower.5ex\hbox{$\sim$}}}\ht1=\grdimen\dp1=0pt
\setbox\simlessbox=\hbox{\raise.5ex\hbox{$<$}\llap
     {\lower.5ex\hbox{$\sim$}}}\ht2=\grdimen\dp2=0pt

\begin{document}

\pagestyle{empty}

\centerline{\Large {\bf Accretion in the Early Outer Solar System}}
\vskip 7ex
\centerline{Scott J. Kenyon}
\centerline{Harvard-Smithsonian Center for Astrophysics}
\centerline{60 Garden Street, Cambridge, MA 02138} 
\centerline{e-mail: skenyon@cfa.harvard.edu}
\vskip 3ex
\centerline{and}
\vskip 3ex
\centerline{Jane X. Luu}
\centerline{Leiden Observatory}
\centerline{PO Box 9513, 2300 RA Leiden, The Netherlands} 
\centerline{e-mail: luu@strw.leidenuniv.nl}
\vskip 7ex
\centerline{to appear in }
\centerline{{\it The Astrophysical Journal}}

\vskip 15ex

\begin{abstract}

We describe calculations of the evolution of an ensemble of small
planetesimals in the outer solar system.  In a solar nebula with a
mass of several times the Minimum Mass Solar Nebula, objects with
radii of 100--1000 km can form on timescales of 10--100 Myr.  Model
luminosity functions derived from these calculations agree with
current observations of bodies beyond the orbit of Neptune (Kuiper
Belt objects).  New surveys with current and planned instruments can
place better constraints on the mass and dynamics of the solar nebula
by measuring the luminosity function at red magnitudes, $\bf m_R \ge
28$.

\end{abstract}

\subjectheadings{solar system: formation -- Kuiper Belt}

\section{INTRODUCTION}
\pagestyle{plain}

Several remarkable discoveries have renewed interest in solar system
formation.  Recent surveys have detected many small icy bodies beyond
the orbit of Neptune (e.g., \cite{jew93}; \cite{wil95}; \cite{jew96}; 
\cite{luu97}; \cite{gla97}; \cite{chi99}).
Assuming a geometric albedo of 4\%, these Kuiper Belt objects (KBOs)
have radii of 50--400 km; the derived size distribution implies a 
significant population of smaller objects.  With semi-major axes 
of 40--50 AU and orbital inclinations of 0\degree--30\degree, the
orbits of known KBOs suggest an annulus of planetesimals formed {\it
in situ} and left over from the planetary formation epoch (\cite{hol93}).
The presumed structure of this annulus resembles the dusty disks recently
discovered around several nearby stars (\cite{smi84}; \cite{aum84};
\cite{jay98}; \cite{koe98}; \cite{gre98}).  
Planets similar to those in our solar system have not
been detected in any of these disks, but direct images and radial
velocity measurements of other nearby stars already imply the existence 
of more than one dozen extra-solar planets of several Jupiter masses
(\cite{lat89}; \cite{mar96}; \cite{coc97};
\cite{noy97}; \cite{del98}; for a review see \cite{mar99}).

These discoveries challenge planet formation theories.  Most theories
presume that planets grow by accretion of small planetesimals in a
gaseous circumstellar disk (\cite{saf69}; \cite{gol73}; see also
\cite{lis93}, \cite{bos97}, and references therein).
Hitherto, numerical studies have focused on the formation of the
prototypical terrestrial and gas giant planets, Earth and Jupiter
(\cite{gre78}, 1984; \cite{nak83}; \cite{oht88}; \cite{wet89}, 1993; 
\cite{bar90}; \cite{rud91}; \cite{pol96}; \cite{wei97}).  If the initial 
disk mass is comparable to the Minimum Mass Solar Nebula\footnote{
The
Minimum Mass Solar Nebula has a surface density $\Sigma = \Sigma_0
(R/R_0)^{-3/2}$, where $\Sigma_0$ is the surface density of solid 
material at $R_0$ = 1 AU.  We adopt $\Sigma_0$ = 45 g cm$^{-2}$ 
(\cite{hay81}; see also \cite{wei77}; \cite{bai94}). This 
definition yields a total mass of solids 
$M_0 \approx$ 100 $M_E$ inside the orbit of Neptune ($R <$ 30 AU)
and $M_0 \approx$ 10 $M_E$ in the inner part of the Kuiper Belt 
($R$ = 32--38 AU), where 1 $M_E = 6 \times 10^{27}$ g is the mass 
of the Earth.}, these calculations often have difficulty producing 
objects similar to the known terrestrial or gas giant planets during 
the estimated disk lifetime of $\sim$ 10--30 Myr (\cite{pol96};
\cite{wei97}).  
This problem is exacerbated in the outer solar system, where 
numerical calculations yield formation times exceeding 100 Myr 
for 500--1000 km radius KBOs (\cite{ste95}, 1996; \cite{st97a},b).
KBOs must form
on shorter timescales in parallel with Neptune.  Otherwise, 
Neptune's gravity increases the velocities of nearby planetesimals, 
including those in the inner Kuiper Belt, on timescales of 
$\sim$ 10 Myr (\cite{mal96}).  This process prevents the 
growth of KBOs with radii exceeding 100--200 km, because 
large velocities hinder agglomeration.

We recently began to consider KBO formation in the outer solar system
using an evolution code that follows planetesimal growth in the
annulus of a circumstellar disk. Initial results indicate that KBOs
can form at 30--50 AU on timescales of 10--100 Myr in disks with 1--3
times the Minimum Mass Solar Nebula when collisional disruption of
planetesimals is unimportant (Kenyon \& Luu 1998; hereafter \cite{kl98}).
Further calculations with an algorithm that includes disruptive 
processes lead to similarly short timescales for a wide range of 
initial conditions (Kenyon \& Luu 1999; hereafter \cite{kl99}).
Here, we briefly summarize these new results, compare the theoretical
model with current observations, and make predictions for comparison
with future observations of KBOs.

\section{MODEL}

Our accretion code is based on the particle-in-a-box method, where
planetesimals are a statistical ensemble of bodies with a distribution 
of horizontal and vertical velocities about Keplerian orbits 
(\cite{saf69}).  We perform calculations for a single annulus of 
width $\Delta a$ centered at a heliocentric distance $a$.
We approximate the continuous distribution of particle masses with 
$i$ discrete batches having particle populations $n_i(t)$ and total masses
$M_i(t)$. The horizontal and vertical velocity dispersions are 
$h_i(t)$ and $v_i(t)$ (\cite{wet93}). The average mass of a 
batch, $m_i(t)$ = $M_i(t) / n_i(t)$, changes with time as collisions 
add and remove bodies from the batch.  This procedure conserves mass 
and provides a statistical method to follow the growth of $\gtrsim$
$ 10^{20}$ small planetesimals into a few planets.  Detailed $n$-body 
calculations confirm the basic features of particle-in-a-box calculations 
for the early stages of planet growth described here (\cite{ida92};
\cite{kok96}).

To evolve the initial size distribution in time, we calculate 
collision rates for the coagulation equation,
determine the outcome of each collision,
and compute velocity changes due to collisions and 
long-range gravitational interactions (see KL99).  
Each two-body collision can produce
(1) merger into a single body with no escaping debris (very low impact velocity),
(2) merger into a single body with escaping debris 
(`cratering'; low impact velocity),
(3) rebound with or without cratering (modest impact velocity), or
(4) catastrophic disruption into numerous smaller bodies 
(high impact velocity).  The collision outcomes depend on the ratio 
of the impact energy $Q_f$ to the disruption energy $Q_d$ of 
two colliding planetesimals (\cite{gre78}; \cite{wet93}; \cite{dav94}).  
Collisions with $Q_f > Q_d$
disrupt planetesimals into many small fragments.  Collisions
with $Q_f < Q_d$ yield a merged planetesimal and some small fragments
if the collision velocity $V_c$ exceeds the minimum velocity for 
cratering $V_f$.  Collisions with $Q_f < Q_d$ and $V_c < V_f$ yield
a merged planetesimal with no cratering debris.  We use an energy-scaling 
formalism to compute $Q_d$ as the sum of the intrinsic tensile 
strength $S_0$ and the gravitational binding energy (\cite{dav85}, 1994).  
The intrinsic strength is the dominant component of $Q_d$ for 
bodies with $r_i \lesssim$ 1~km; gravitational binding dominates 
$S_0$ for larger bodies.  

For each collision, a velocity evolution algorithm distributes the kinetic
energy among the resulting bodies and then accounts for collisional 
damping, kinetic energy transfer during elastic collisions (``dynamical 
friction''), angular momentum transfer during elastic collisions 
(``viscous stirring''), and gas drag (\cite{hor85}; see also \cite{wet93};
\cite{kl98}).  Dynamical friction tries to enforce equipartition 
of kinetic energy between mass batches; 
viscous stirring increases the velocities of all bodies.  Gas drag 
removes objects from the annulus and reduces the velocities of small 
objects which are well coupled to gas in the disk. 

We tested the code against analytical solutions and published numerical
results (\cite{kl98}, \cite{kl99}).  We reproduced previous calculations 
for accretion at 1 AU (\cite{wet93}) and collisional disruption of
pre-existing large KBOs at 40 AU (\cite{dav97}). Our calculations match 
analytical solutions 
well when the mass spacing between successive batches, $\delta$ =
$m_{i+1}/m_i$ = 1.1--1.4.  Numerical solutions lag the analytic
results by $\sim$ 10\% when $\delta$ = 1.4--2.  The timescale to
produce objects of a given size increases with $\delta$, because
poorer resolution prevents growth of large objects (see \cite{kl98}).

Table 1 lists basic input parameters.  The input cumulative size 
distribution $N_C$ has the form $N_C \propto r_i^{q_0}$, with 
initial radii $r_i$ = 1--80 m.  The total mass in the annulus 
is $M_0$; $M_0 \approx$ 10 $M_E$ for a Minimum Mass Solar Nebula.
All batches start with the same initial velocity.  We tested a 
range of initial velocities corresponding to initial eccentricities 
of $e_0 = 10^{-4}$ to $10^{-2}$, as is expected for planetesimals 
in the early solar nebula (\cite{mal95}).  The adopted mass density, 
$\rho_0$ = 1.5 g cm$^{-3}$, is appropriate for icy bodies with a 
small rocky component.  The fragmentation parameters -- $V_f$, 
$S_0$, $Q_c$, $f_{KE}$, $c_1$, and $c_2$ -- are adopted from 
earlier work.  KL99 describe these parameters in more detail.

To provide observational constraints on the models, we note 
that the known Kuiper Belt population contains at least 
one body with a radius of $\sim$ 1000 km  (Pluto), and 
$\sim 10^5$ KBOs with radii $r_i \gtrsim$ 50 km between 30--50 AU.  
The cumulative size distribution of known KBOs can be fitted 
with $N_C \propto r_i^{q_{obs}}$, with $q_{obs} = 3 \pm 0.5$ 
(\cite{jew98}; see also \cite{chi99}).  Successful models 
should reproduce these 
observations on timescales comparable to 
(a) the estimated lifetimes of the solar nebula and gaseous disks 
surrounding nearby young stars, $\lesssim 10^7$ yr (\cite{rus96}; 
\cite{har98}) and (b) the formation timescale for Neptune, 
$\lesssim 10^8$ yr (\cite{lis96}).

\section{NUMERICAL RESULTS}

We separate the growth of KBOs into three regimes.  Early in the 
evolution, frequent collisions damp the velocity dispersions of small 
bodies.  These bodies slowly grow into 1~km objects on a timescale 
that is approximated by $\tau_{\rm 1~km} \approx$ 8 Myr 
$(M_0/10 M_E)~(e_0/10^{-3})^{0.65}$.
This linear growth phase ends when the gravitational range of the largest
objects exceeds their geometric cross-section.  This ``gravitational focusing''
enhances the collision rate by factors of 10--1000.  The largest objects 
then begin a period of ``runaway growth'', when their radii grow from 
$\sim$ 1~km to $\gtrsim$ 100~km in several Myr.  During this phase, 
dynamical friction and viscous stirring increase the velocity dispersions 
of the smallest bodies from $\sim$ 1~m~s$^{-1}$ up to $\sim$ 40~m~s$^{-1}$.
This velocity evolution reduces gravitational focusing factors and ends 
runaway growth.  The largest objects then grow slowly to 1000+ km sizes 
on timescales that again depend on $M_0$ and $e_0$.  Column (5) in 
Table 2 lists timescales to form Pluto-size objects $\tau_P$ as 
a function of the input parameters $M_0$, $\delta$, $e_0$, and $q_0$.

Fig. 1 shows cumulative size distributions for a model with $M_0 =
10~M_E$, $q_0$ = 3, and $S_0 = 2 \times 10^6$ erg g$^{-1}$.  The
shapes of these curves depend on two competing physical processes: 
(1) growth by mergers and (2) erosion by high velocity collisions.  
In this example, collisions result in growth because the velocity
dispersion is less than the catastrophic disruption threshold.
However, the collision velocity exceeds the cratering threshold $V_f$.
Cratering adds debris to all low mass batches. Gas drag removes
material from low mass batches ($r_i \lesssim$ 10~m), but is
ineffective at removing larger objects.  The size distribution thus
becomes shallower at small masses.  At large masses, mergers produce 
a group of growing planetesimals with a steep size distribution.
Once gravitational focusing becomes effective, the largest of these 
objects `run away' from the rest of the ensemble to produce a smooth 
power law with a maximum radius $r_{max}$.  As the evolution proceeds,
$r_{max}$ increases but the slope of the smooth power law remains nearly 
constant.

The main features of these results depend little on the input parameters.
All calculations produce two cumulative power law size distributions 
connected by a transition region having an `excess' of planetesimals 
(the ``bump'' in the curves in Fig. 1).  The characteristic radius
of this transition region increases from 0.3~km at $e_0 = 10^{-4}$ to
3~km at $e_0 = 10^{-2}$. If fitted with a power law of the form $N_C
\propto r_i^{-q_f}$ at small masses, the cumulative size distribution
follows the predicted limit for collisional evolution, $q_f = 2.5$
(\cite{doh69}). We perform least-square fits to obtain $q_f$ at larger
masses; column (6) of Table 2 lists derived values for $q_f$ along 
with the 1$\sigma$ error. Column (7) lists the radius range for 
each fit.  The results are surprisingly independent of the input
parameters.  We find the small range $q_f$ = 2.75--3.25 for calculations
with $M_0$ = 1--100 $M_E$, $e_0 = 10^{-4}$ to $10^{-2}$, $q_0$ =
1.5--4.5, and $S_0$ = 10 erg g$^{-1}$ to $3 \times 10^6$ erg g$^{-1}$.
This model result is consistent with the observed slope, $q_{obs} = 3
\pm 0.5$ (e.g., Jewitt {\it et al.} 1998).

\section{COMPARISONS WITH OBSERVATIONS}

As shown in Table 2, several calculations meet the success criteria defined in $\S 2$.
Annuli with $M_0 \gtrsim 10 M_E$ produce Pluto-sized objects on 
short timescales, $\tau_P \lesssim$ 50 Myr (for $e_0 \lesssim 10^{-3}$).  
Models with smaller initial masses or larger initial eccentricities 
form Plutos on longer timescales, $\tau_P \gtrsim$ 50 Myr.  Plausible 
ranges of other input parameters -- such as $q_0$, $S_0$, and 
$f_{KE}$ -- yield $\pm$20\% variations about these timescales.
The results are insensitive to $V_f$ and other collision parameters
(\cite{kl99}).

The crosses in Figure 1 compare our calculations directly with 
several observational constraints.  The cross at $r_i$ = 50~km 
indicates the number of KBOs with $r_i \gtrsim$ 50 km estimated 
from recent ground-based surveys (\cite{jew98}; see also \cite{chi99}); 
the one at 
$r_i$ = 10~km shows limits derived from a single, controversial 
measurement with {\it Hubble Space Telescope} ({\it HST}; 
\cite{coc95}, 1998; \cite{bro97}).  The third cross plots limits
at $r_i$ = 1~km derived from theoretical attempts to explain 
the frequency of short-period comets from the Kuiper Belt 
(\cite{dav97}; \cite{dun97}; \cite{lev97}).  Our predictions 
agree with ground-based surveys at 50~km and theoretical 
limits at 1~km, but fall a factor of $\sim$ 10 short of 
the {\it HST} measurement at 10~km.

To compare with observations in more detail, we predict the luminosity
function (LF) of KBOs directly from the computed number distribution.
We use a Monte Carlo calculation of objects selected randomly from the
cumulative size distribution $N_C$.  We assign each object a distance
from the Sun $d_{\sun}$ and a random phase angle $\beta$ between the
line-of-sight from the Earth to the object and the line-of-sight from
the Sun to the object.  This phase angle lies between 0\degree~and a
maximum phase angle that is distance-dependent. The distance of the
object from the Earth is then $d_E = d_{\sun} {\rm cos} \beta - (1 +
d_{\sun}^2 ({\rm cos}^2 \beta - 1))^{1/2}$. We derive the red
magnitude of this object from a two parameter magnitude relation for
asteroids, $m_{R,KBO} = R_0 + 2.5~{\rm log}~(t_1/t_2) - 5~{\rm log}~
r_{KBO}$, where $R_0$ is the zero point of the magnitude scale,
$r_{KBO}$ is the radius of the KBO, $t_1 = 2 d_{\sun} d_E$, and $t_2 =
\omega ( (1-g) \phi_1 + g \phi_2)$ (\cite{bow89}).  In this last
expression, $\omega$ is the albedo, and $g$ is the slope parameter;
$\phi_1$ and $\phi_2$ are phase functions that describe the visibility
of the illuminated hemisphere of the object as a function of $\beta$.
We adopt standard values, $\omega = 0.04$ and $g = 0.15$, appropriate
for comet nuclei (\cite{jew98}). The zero point $R_0$ is the apparent
red magnitude of the Sun, $m_{R,\sun}$ = $-$27.11, with a correction for 
the V--R color of a KBO, $R_0$ = $m_{R,\sun}$ + $\delta$(V--R)$_{KBO}$.  
Observations suggest that KBOs have colors that range from roughly 
$-0.1$ to 0.3 mag redder than the Sun. We treat this uncertainty by 
allowing the color to vary randomly in this range.

The important parameters in the model LF are the distributions of
input sizes (derived from the accretion calculations), distances,
and orbital parameters.  We assume KBOs are evenly distributed 
between ``Plutinos,'' objects in 3:2 orbital resonance with Neptune 
having semimajor axes of 39.4$\pm$0.2 AU, and ``classical'' 
KBOs with semimajor axes between 42--50 AU.  The distance parameters 
are set by observations (\cite{jew98}).  This distance distribution 
is different from the 32--38 AU adopted for the coagulation 
calculations. Several tests show that accretion results at 
42--50 AU are identical to those at 32--38 AU, except that the 
timescale to produce Pluto-sized objects is 50\%--100\% longer.  
To compute the model LF from the Monte Carlo magnitude distribution 
of KBOs, we scale the mass in the 32--38 AU annulus to match the 
mass in a 42--50 AU annulus, add in an equal number of Plutinos, 
and divide by the sky area.  The distribution of KBO orbital 
parameters is poorly known. We adopt circular orbits to derive 
magnitudes; the LF is insensitive to other choices.  We assume 
orbital inclinations of $i =$ 0\degree~to 5\degree~to compute 
the sky area, which is a compromise between the $i \approx$ 
0\degree--5\degree~of classical KBOs and the $i \approx$
10\degree--30\degree~of Plutinos. The model LFs scale inversely 
with sin $i$.

Figure 2 compares several models with the observed LF.  The left panel
shows models with $e_0$ = $10^{-3}$ and different masses; the right
panel shows models with the mass of a Minimum Mass Solar Nebula and 
different $e_0$.  Model LFs with the Minimum Mass and any $e_0$ agree 
with current observations.  The good agreement of all models at $m_R \le$ 20, 
where the uncertainties are largest, depends on the assumed maximum 
radius in the model distribution.  We picked 1000 km for convenience.  
Model LFs for $m_R \ge$ 20 are independent of this choice.

To quantify the comparison between models and observations,
we fit model LFs to log $\Sigma (m_R)$ = $\alpha (m_R - m_0)$ 
over $ 20.5 \le m_R \le 26.5 $. Table 3 lists the fitted $\alpha$ 
and $m_0$ as a function of the mass in classical KBOs (in units 
of the Minimum Mass Solar Nebula), $e_0$, the inner annulus 
boundary $R_{in}$, and the outer annulus boundary $R_{out}$.
The small range in $\alpha$ for model LFs agrees with 
published values derived from observations\footnote{Gladman \etal 
(1998) report $\alpha = 0.76_{-0.11}^{+0.10}$ and $m_0 = 
23.4_{-0.18}^{+0.20}$ from a maximum likelihood analysis of 
previous surveys with magnitude limits, 20 $\le m_R \le$ 28.  
Surface densities in their Table 3 yield $\alpha \approx 0.6$ 
and $m_0 \approx $ 22.4.  Jewitt \etal (1998; see also Luu \&
Jewitt 1998) quote $\alpha = 0.54 \pm 0.04$ and $m_0 = 23.2 
\pm 0.10$ for 20 $\le m_R \le$ 26. Chiang \& Brown (1999) prefer
$\alpha = 0.52 \pm 0.02$ and $m_0$ = 23.5 for  20 $\le m_R \le$ 27;
they note that the slope depends on which survey data are used in
the fit.}. The model $\alpha$ is independent 
of the relative numbers of Plutinos and classical KBOs, and the 
distance distribution of classical KBOs.  The observed zero-point 
of the LF, $m_0 \approx$ 23.2--23.5, favors models with masses 
comparable to the Minimum Mass Solar Nebula and any initial 
eccentricity.  These data rule out models with $\le 30\% $
of the Minimum Mass at the 3$\sigma$ level.  Smaller Plutino
fractions require larger masses: if Plutinos are 10\%--25\% 
of the total KBO population, as indicated by recent 
observations (\cite{jew98}), the needed mass is 2--4 times 
the Minimum Mass.  

There are two main uncertainties in comparing our model LFs 
with the data, the evolution of the KBO LF with time and the 
current orbital parameters of KBOs.  The initial mass in KBOs 
was larger than implied by a direct comparison between the data 
and model LFs, because large velocity collisions and dynamical 
encounters with Neptune have eroded the Kuiper Belt over time 
(\cite{hol93}; \cite{dav97}; see also \cite{lev93}; \cite{dun95}).
Erosion from high velocity collisions probably does not change 
the slope of the LF significantly.  Massive KBOs with $r_i \gtrsim$ 
50 km ($m_R \lesssim$ 26--27) are probably safe from collisional 
disruption (\cite{dav97}).  Disruption of smaller bodies depends on 
the unknown bulk properties and the poorly known orbital parameters
of KBOs.  These uncertainties are not important for comparisons 
of models and observations for $m_R \lesssim$ 26--27, but can
bias future comparisons at fainter magnitude limits.  
Gravitational perturbations from Neptune should affect all 
KBO masses equally and simply reduce the total mass in KBOs 
with time (\cite{hol93}).  Despite the uncertainty in the 
total amount of mass lost from the Kuiper Belt, we are 
encouraged that the mass needed to explain current observations 
of KBOs is at least the Minimum Mass Solar Nebula.  Future 
calculations will allow us to place better constraints on 
the initial mass in the Kuiper Belt.

The uncertain distribution of KBO orbital parameters also affects the 
initial mass estimates.  Our assumption of KBOs uniformly distributed 
in distance $d_{\sun}$, orbital eccentricity $e$, and inclination
$i$ is probably incorrect for Plutinos in specific orbital 
resonances with Neptune.  Larger adopted volumes for current 
Plutinos require larger initial disk masses in the Kuiper Belt.
A uniform distribution is probably reasonable for classical KBOs,
but the observed range in $d_{\sun}$ and $i$ is not well-known.
Allowing classical KBOs to occupy a larger range in semi-major
axis reduces our mass estimates; a larger range in sin $i$ 
increases our mass estimates.  We suspect that the uncertainties
currently are a factor of $\sim$ 2--3.  Future large area surveys 
will provide better knowledge of KBO orbital parameters and allow 
more accurate models for the observed LF.

In addition to the reasonably good fit for $ 20.5 \le m_R \le 26.5 $,
our calculations predict 1--5 `Plutos' with $m_R \le$ 20 over the
entire sky.  This number is uncertain, because we do not understand
completely the mechanism that ends accretion and sets the maximum size
of KBOs.  Our calculations indicate that planetary accretion at 35--50 AU
is self-limiting: once objects reach radii of $\sim$ 1000 km, they
stir up smaller bodies sufficiently to limit additional growth.
The formation of nearby Neptune should have also limited the growth
of the largest bodies (\cite{mor97}).  Better
constraints on the radial distribution of 500+ km KBOs would test the
relative importance of these two mechanisms.

Observations at fainter magnitude limits will provide additional
constraints on KBO formation.  Imaging data acquired at the Keck and
Palomar telescopes detect KBOs with $m_R \approx$ 25--26.5, where
models with $e_0 \gtrsim 10^{-2}$ predict the LF to rise sharply.  The
apparent lack of a significant upturn in the LF at $m_R \le$ 25
implies $e_0 \lesssim$ a few $\times~10^{-2}$.  In contrast, the
current limit on the KBO population at $m_R \ge$ 28 implies a
substantial population of 10 km radius KBOs which is inconsistent with
our calculations.  Deeper ground-based surveys or new {\it HST} data
could resolve the controversy surrounding this observation and place
better constraints on $e_0$.  Finally, the proposed {\it Next
Generation Space Telescope} ({\it NGST}) will probe the size
distribution of 1~km radius KBOs where models with $e_0 \approx
10^{-3}$ predict a sharp upturn in the observed LF.  If such small
bodies can survive for the age of the solar system, {\it NGST}
observations would provide important constraints on the initial mass
and dynamics of the outer solar system.

\vskip 6ex

We thank B. Bromley for making it possible to run our code 
on the JPL Cray T3D `Cosmos' and the HP Exemplar `Neptune' 
and for a generous allotment of computer time through 
funding from the NASA Offices of Mission to Planet Earth, 
Aeronautics, and Space Science.  Comments from F. Franklin, 
M. Geller, and M. Holman greatly improved our presentation.  

\vfill
\eject

\vfill
\eject

\pagestyle{empty}

\singlespace
\begin{center}
\begin{tabular}[t]{l c c l}
\multicolumn{4}{c}{{\sc Table 1.} Basic Model Parameters} \\
\\
\tableline
\tableline
Parameter & Symbol & \hspace{5mm} & Value \\
\tableline
Width of annulus & $\delta a$ && ~~6 AU \\
Initial velocity & $V_0$      && ~~0.45--45 m s$^{-1}$ \\
Particle mass density & $\rho_0$ && ~~1.5 g cm$^{-3}$ \\
Relative gas velocity & $\eta$ && ~~30 m s$^{-1}$ \\
Time step        & $\delta t$ && ~~1--250 yr \\
Number of mass bins & $N$     && ~~64--256 \\
Mass spacing of bins& $\delta$ && ~~1.25-2.0 \\
Minimum velocity for cratering & $V_f$ && ~~1 cm s$^{-1}$ \\
Impact strength & $S_0$ && ~~$2 \times 10^6$ erg g$^{-1}$ \\
Crushing energy & $Q_c$ && ~~$5 \times 10^{7}$ erg g$^{-1}$ \\
Fraction of KE in ejecta & $f_{KE}$ && ~~0.05 \\
Coefficient of restitution & $c_1$ && ~~$10^{-2}$ \\
Coefficient of restitution & $c_2$ && ~~$10^{-3}$ \\
\tableline
\end{tabular}
\end{center}

\vfill
\eject

\begin{center}
\begin{tabular}[t]{lcccccc}
\multicolumn{7}{c}{{\sc Table 2.} Model Results at 32--38 AU} \\
\\
\tableline
\tableline
$M_0 (M_E)$~ & ~~~$\delta$~~~ & ~~~$e_0$~~~ & ~~~$q_0$~~~ & ~~~$\tau_P$ (Myr)~~~ & $q_f$ & ~~~Range (km)~~~ \\
\tableline
1  & 1.4 & $10^{-4}$ &$-3.0$ & 448 & $-2.86\pm0.04$ & 2--300 \\
10 & 1.4 & $10^{-4}$ &$-3.0$ &  ~20 & $-2.90\pm0.02$ & 1--930 \\
\\
1  & 1.4 & $10^{-3}$ &$-3.0$ & 893 & $-2.76\pm0.05$ & 6--400 \\
3  & 1.4 & $10^{-3}$ &$-3.0$ & 184 & $-2.78\pm0.03$ & 5--600 \\
10 & 1.4 & $10^{-3}$ &$-3.0$ & ~37 & $-2.91\pm0.03$ & 6--800 \\
30 & 1.4 & $10^{-3}$ &$-3.0$ & ~10 & $-3.02\pm0.03$ & 6--930 \\
100& 1.4 & $10^{-3}$ &$-3.0$ & ~~3 & $-2.97\pm0.03$ & 7--600 \\
\\
10  & 1.4 & $10^{-2}$ &$-3.0$& 428 & $-3.15\pm0.10$ & 50--700 \\
100 & 1.4 & $10^{-2}$ &$-3.0$& ~25 & $-3.23\pm0.06$ & 20--600 \\
\\
10  & 1.25 & $10^{-3}$ &$-1.5$& ~40 & $-2.97\pm0.02$ & 7--700 \\
10  & 1.25 & $10^{-3}$ &$-3.0$& ~35 & $-3.03\pm0.03$ & 9--650 \\
10  & 1.25 & $10^{-3}$ &$-4.5$& ~30 & $-2.90\pm0.02$ & 4--750 \\
\tableline
\end{tabular}
\end{center}

\vfill
\eject

\begin{center}
\begin{tabular}[t]{ccccccc}
\multicolumn{7}{c}{{\sc Table 3.} Luminosity Function Parameters} \\
\\
\tableline
\tableline
$M_0/M_{MMSN}$& $ e_0$ & $R_{in}$ (AU) & $R_{out}$ (AU) & \hspace{5mm} & $\alpha$ & $m_0$ \\ 
\tableline
0.3 & $10^{-3}$ & 42 & 50 & & 0.56 $\pm$ 0.01 & 24.03 $\pm$ 0.16 \\
1.0 & $10^{-4}$ & 42 & 50 && 0.58 $\pm$ 0.01 & 23.36 $\pm$ 0.12 \\
1.0 & $10^{-3}$ & 42 & 50 && 0.57 $\pm$ 0.02 & 23.16 $\pm$ 0.15 \\
1.0 & $10^{-2}$ & 42 & 50 && 0.60 $\pm$ 0.01 & 23.53 $\pm$ 0.18 \\
3.0 & $10^{-3}$ & 42 & 50 && 0.58 $\pm$ 0.01 & 22.42 $\pm$ 0.17 \\
\\
0.3 & $10^{-3}$ & 42 & 60 && 0.56 $\pm$ 0.01 & 23.50 $\pm$ 0.13 \\
1.0 & $10^{-4}$ & 42 & 60 && 0.57 $\pm$ 0.01 & 22.85 $\pm$ 0.11 \\
1.0 & $10^{-3}$ & 42 & 60 && 0.56 $\pm$ 0.02 & 22.63 $\pm$ 0.13 \\
1.0 & $10^{-2}$ & 42 & 60 && 0.63 $\pm$ 0.02 & 23.31 $\pm$ 0.19 \\
3.0 & $10^{-3}$ & 42 & 60 && 0.58 $\pm$ 0.01 & 21.92 $\pm$ 0.17 \\
\tableline
\end{tabular}
\end{center}

\hskip 5ex
\epsfxsize=7.5in
\epsffile{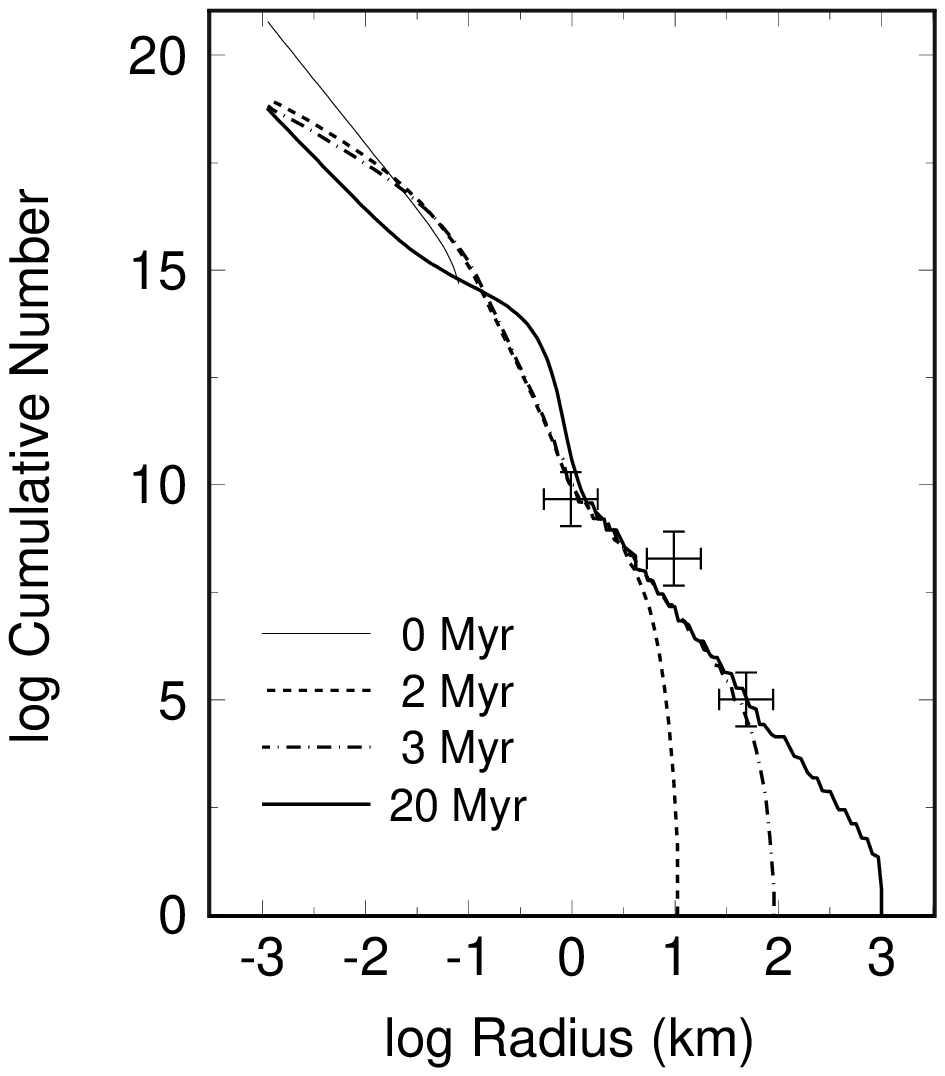}
\figcaption{Cumulative size distributions as a function
of time for a model with $M_0 = 10 M_E$ and $e_0 = 10^{-4}$.  
The evolution time for each curve is listed in the legend. 
Crosses indicate observational and theoretical constraints 
on the size distribution at radii of 50 km, 10 km, and 1 km
as described in the text.}

\hskip -10ex
\epsfxsize=6.750in
\epsffile{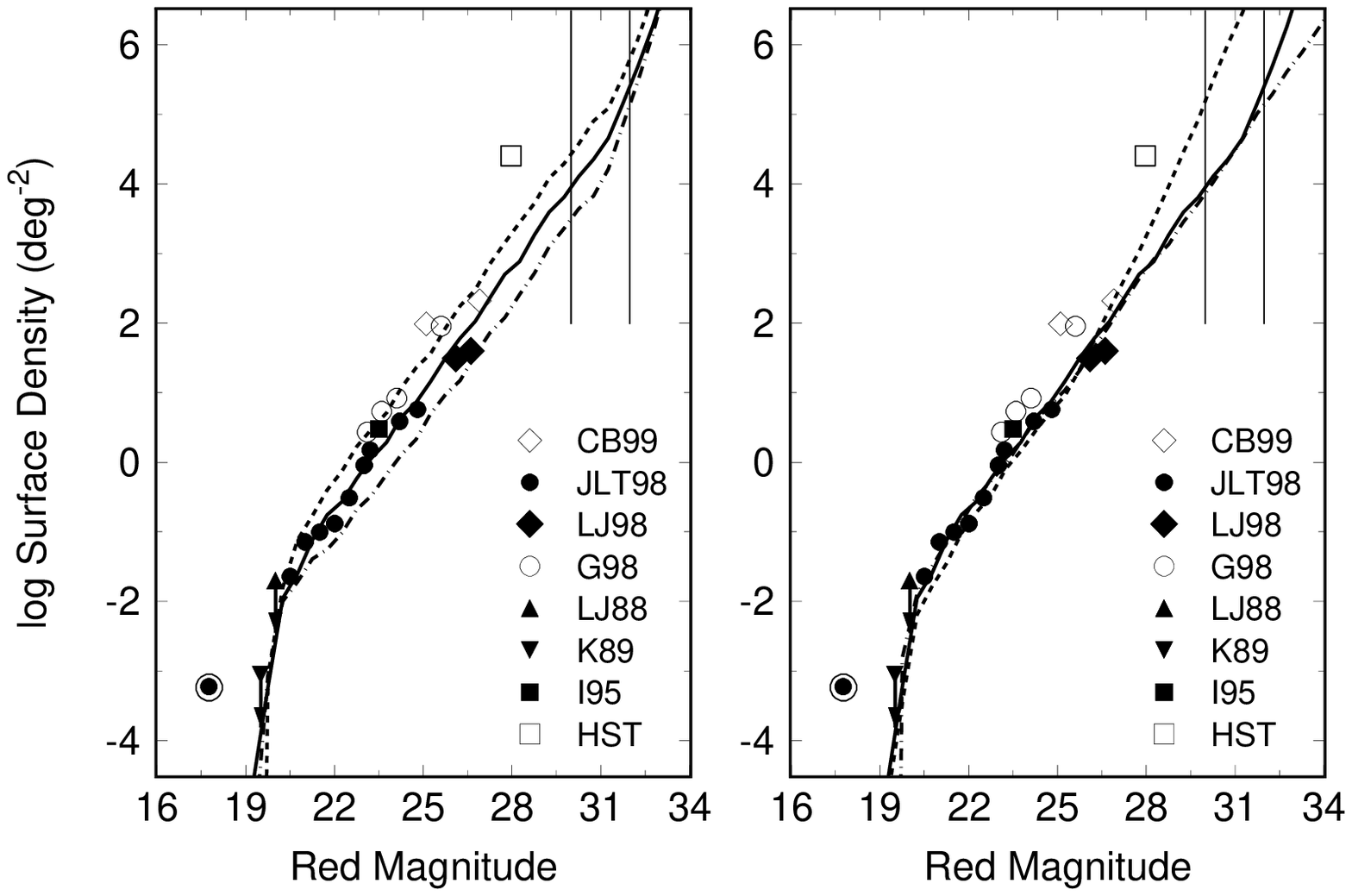}
\figcaption{Comparison of model luminosity functions of KBOs
with observations.  Data are as indicated in the legend of each panel.
The open circle with the central dot is the position of Pluto for an
adopted albedo of 4\%; other observations are from 
Cochran \etal (1998; HST), Irwin \etal (1995; I95), 
Kowal 1989 (1989; K89), Luu \& Jewitt (1988; LJ88),
Gladman \etal (1998; G98), Luu \& Jewitt (1998; LJ98), 
Jewitt \etal (1998; JLT98) and Chiang \& Brown (1999; CB99).  
Error bars for each datum -- typically a factor of 2--3 -- and 
the upper limit from Levison \& Duncan (1990) are not shown 
for clarity. The lines plot luminosity functions for models 
with (a) left panel: $e_0 = 10^{-3}$ and $M_0
\approx$ 0.3 (dot-dashed), 1.0 (solid), and 3.0 (dashed) times the
Minimum Mass Solar Nebula and (b) right panel: a Minimum Mass Solar
Nebula with $e_0 = 10^{-2}$ (dashed), $e_0 = 10^{-3}$ (solid), and
$e_0 = 10^{-4}$ (dot-dashed).  A Minimum Mass Solar Nebula has
$M_0 \approx 12~M_E$ within R = 42--50 AU.  The pair of vertical solid 
lines indicates the planned magnitude range accessible to {\it NGST}.}
\end{document}